\documentclass[a4paper,12pt]{article}
\usepackage{epsfig}
\usepackage{citesort}
\usepackage{psfrag}
\date{\today}
 \normalsize

\def\be{\begin{equation}}
\def\ee{\end{equation}}
\def\bea{\begin{eqnarray}}
\def\eea{\end{eqnarray}}

\def\lsim{\raise0.3ex\hbox{$\;<$\kern-0.75em\raise-1.1ex\hbox{$\sim\;$}}}
\def\gsim{\raise0.3ex\hbox{$\;>$\kern-0.75em\raise-1.1ex\hbox{$\sim\;$}}}
\def\Frac#1#2{\frac{\displaystyle{#1}}{\displaystyle{#2}}}
\def\susy{\mbox{\tiny SUSY}}
\def\sm{\mbox{\tiny SM}}

\def\im{{\mbox{Im}}}
\def\hc{{\mbox{H.c.}}}
\def\ecm{{\mbox{{\it{e}}\,cm}}}

\def\dd#1#2{{\left(\delta^d_{#1}\right)_{#2}}}

\def\ep{\eta^{\prime}}
\begin{document}
\renewcommand{\thefootnote}{\fnsymbol{footnote}}
\rightline{IPPP/04/87} \rightline{DCPT/04/174} \vspace{.3cm} {\Large
\begin{center}
{\bf EDM constraints and CP asymmetries of $B$ processes in
supersymmetric models}
\end{center}}
\vspace{.3cm}
\begin{center}
S. Abel$^{1}$ and S. Khalil$^{1,2}$\\
\vspace{.3cm} $^1$ \emph{IPPP, University of Durham, South Rd.,
Durham DH1 3LE, U.K.}\\
$^2$ \emph{Department of Mathematics, German University in Cairo, New Cairo city,
El Tagamoa El Khames, Egypt.}\\
\end{center}
\begin{center}
\small{\bf Abstract}\\[3mm]
\end{center}
\begin{minipage}[h]{14.0cm}
We demonstrate that electric dipole moments (EDMs)
strongly constrain possible SUSY contributions to the CP
asymmetries of $B$ processes; $LL$ and/or $RR$ flavour
mixings between second and third generations are severely
restricted by the experimental
limit on the mercury EDM, and so therefore are their
possible contributions to the CP asymmetries of $B\to \phi K$ and
$B\to \eta' K$. We find that SUSY models with dominant $LR$ and
$RL$ mixing through non-universal $A$-terms is the only
way to accommodate the apparent deviation of CP asymmetries
from those expected in the Standard Model
without conflicting with the EDM bounds or with any other
experimental results.
\end{minipage}
\section{Introduction}
The most recent results of BaBar and Belle collaborations
\cite{babar-belle} on the mixing-induced asymmetries of $B\to \phi
K$ and $B\to \eta' K$ indicate possible deviation from the
Standard Model (SM) expectations. The Belle experimental values of
these asymmetries are given by \bea S_{\phi K} &=& 0.06 \pm 0.33
\pm 0.09, \\
S_{\eta' K} &=& 0.65 \pm 0.18 \pm 0.04. \eea The BaBar
experimental values are \bea S_{\phi K} &=& 0.50 \pm
0.25^{+0.07}_{-0.04}, \\
S_{\eta' K} &=& 0.27 \pm 0.14 \pm 0.03. \eea Comparison
with the world average CP asymmetry $S_{J/\psi
K}=0.726\pm0.03$ shows that the average CP asymmetry of all
$b\to s$ penguin modes from the Belle results is
$0.43^{+0.12}_{-0.11}$, which is $2.4 \sigma$ away from the SM
result, and from the BaBar result is $0.42\pm 0.10$,
a $2.7 \sigma$ deviation.

Supersymmetry (SUSY) is one of the most popular candidates for
physics beyond the SM, and a natural place to look for explanations
of such deviation. Indeed in SUSY models there are many new
sources of CP violation besides the CKM phase. However
stringent constraints on these phases are usually obtained from
the experimental bounds on the electric dipole moment (EDM) of the
neutron, electron and mercury atom. Because of this it is it a
challenge for SUSY models to give a new source of CP violation
that can explain the possible discrepancy between CP
asymmetry measurements and the expected SM results, whilst at the
same time avoiding the overproduction of EDMs.

It is known \cite{emidio,susy-cont,KK} that SUSY models
with a large squark mixing and order one phase between the second
and third generations can accommodate the CP asymmetry results via
gluino exchange. The squark mixings can be classified,
according to the chiralities of their quark superpartners,
into left-handed or right-handed  (L or R) squark mixing.
The left-handed mixings for
the down-squark are given by the mass insertions
$(\delta^d_{LL})_{ij}$ and $(\delta^d_{LR})_{ij}$, and the
right-handed mixings by $(\delta^d_{RL})_{ij}$ and
$(\delta^d_{RR})_{ij}$. It is remarkable that in order
simultaneously to satisfy the measurements of $S_{\phi K}$ and $S_{\eta' K}$
and explain the deviation between them, both left- and
right-handed contributions have to be involved. This is because
the left- and right-handed contributions have an opposite sign due to the
different parity in the final states of $B \to \phi K$ and $B\to
\eta' K$ \cite{KK}.

In this paper we argue that a large flavour mixing between the second
and third generation via $(\delta^d_{LL})_{23}$ and
$(\delta^d_{RR})_{23}$ leads to a large
$(\delta^d_{LR})_{22}$, which produces a large
strange EDM and consequently overproduces
neutron EDM (assuming the Parton model) and mercury EDM.
We will show that, taking EDM constraints into account,
the possible solution of the $S_{\phi K}$ and $S_{\eta' K}$
discrepancy based on $(\delta^d_{LL})_{23}$ and
$(\delta^d_{RR})_{23}$ is disfavoured. This leaves the scenario with
large mass insertions $(\delta^d_{LR})_{23}$ and
$(\delta^d_{RL})_{23}$ (due to non-universal trilinear $A$-terms)
as the only possible consistent solution.

This paper is organized as follows. In section 2 we introduce the
supersymmetric contributions to the strange quark EDM which could
be enhanced by large mixing between the second and third
generation and leads to a large Hg EDM. Section 3 is devoted to
imposing the EDM constraints on the SUSY phases in a model
independent analysis, and the impact of these constraints on the
SUSY contribution to the CP asymmetries of $B\to \phi K$ and $B\to
\eta' K$. In section 4 we give numerical results and show
correlations among the Hg EDM and the CP asymmetries of B-decays.
Our conclusions are given in section 5.

%
\section{Supersymmetric contributions to strange quark EDM}
As mentioned in the introduction, SUSY models have several
possible sources of CP violation in addition to the CKM
phase. These CP phases can have important implications for
CP violating phenomenology. In particular they can induce
large EDMs of quarks and
leptons at the one-loop level that far
exceed the experimental limits, and
stringent constraints on SUSY CP phases are found
\cite{abel}. The most recent measurements for the neutron
($d_n$) and mercury ($d_{H_g}$) EDMs lead to the following limits: \bea
d_n &=& 6.3
\times 10^{-26} \ecm , \\
d_{H_g} &=& 2.1 \times 10^{-28} \ecm~. \eea The neutron EDM
receives contributions of different sources and the predicted value
in any particular model depends quite strongly on the particular
model of the neutron used for the calculation. Because of this it is
worth briefly summarizing the calculation.

The major contributions to the EDMs come from
electric and chromoelectric dipole operators and the Weinberg
three-gluon operator: \be {\mathcal L} = -\frac{i}{2} d^E_q
\bar{q} \sigma_{\mu\nu} \gamma_5 q F^{\mu\nu} - \frac{i}{2} d^C_q
\bar{q} \sigma_{\mu\nu} \gamma_5 T^a q G^{a\mu\nu} -\frac{1}{6}
d^G f_{abc} G_{a\mu\rho} G^{\rho}_{b\nu} G_{c\lambda \sigma}
\epsilon^{\mu\nu\lambda\sigma}. \ee In order to evaluate the neutron EDM,
one needs to make some assumptions about the internal
structure of the neutron. The models can be classified as follows; \\

\hspace{-0.65cm}\underline{1- The chiral quark model} \\

In this
model the neutron EDM is related to the EDMs of the valence
quarks; \be d_n = \frac{4}{3} d_d - \frac{1}{3} d_u. \ee The quark
EDMs are given by \be d_q = \eta^E d_q^E + \eta^C \frac{e}{4 \pi}
d_q^C + \eta^G \frac{e \Lambda}{4\pi} d^G, \ee where the QCD
correction factors are given by $\eta^E= 1.53$, $\eta^C\simeq
\eta^G\simeq 3.4$ and where
$\Lambda \simeq 1.19$ GeV is the chiral symmetry breaking scale.\\

\hspace{-0.65cm}\underline{2- The parton quark model} \\

Here one assumes that the quark contributions to neutron EDM are
weighted by the factor $\Delta_q$ defined as $\langle n \vert
\frac{1}{2}\bar{q} \gamma_{\mu} \gamma_5 q \vert n \rangle =
\Delta_q S_{\mu}$, where $S_{\mu}$ is the neutron spin; \be d_n =
\eta^E \left( \Delta_d d_d^E + \Delta_u d_u^E + \Delta_s d_s^E
\right), \ee where the individual quark contributions are given in
terms of the gluino, chargiono and neutralino contributions \be
d_q = d_q^{\tilde{g}} + d_q^{\tilde{\chi}^+} +
d_q^{\tilde{\chi}^0}. \ee The following values for $\Delta_q$ are
usually used: $\Delta_d=0.746$, $\Delta_d=-0.508$, and
$\Delta_s=-0.226$. The main difference between the parton quark
model and the chiral quark model is the large strange quark
contribution in parton model. Also in this model, the relevant
contribution is only due to the electric EDM of the quarks in
contrast with the chiral quark model where the chromoelectric and
three-gluon operators
contribute as well.\\

\hspace{-0.65cm}\underline{3- QCD sum rules}\\

The QCD sum rules analysis of ref.\cite{pospelov} leads to the following
relation between the neutron EDM and the electric EDMs and
chromoelectric EDMs of $u$ and $d$ quarks: \be d_n = 0.7 (d^E_d -
0.25 d^E_u) + 0.55 e (d^C_d + 0.5 d^C_u), \ee where the value of
quark vacuum condensate $\langle\bar{q} q\rangle =(225 \mathrm{GeV})^3$ has
been used. It can be seen from the above equation that the
QCD sum rules cannot
incorporate the effect of the strange quark in the neutron EDM.\\

\hspace{-0.65cm}\underline{4- The Chiral Lagrangian approach}\\

In ref.\cite{hisano}, the chiral lagrangian approach was
adopted to try to incorporate the strange quark chromoelectric EDM
contribution to the neutron EDM. This analysis leads to the
following result for the neutron EDM in terms of the quark
chromoelectric EDM: \be d_n = (1.6~ d^C_u + 1.3~ d^C_u + 0.26~
d^C_s) \ecm~. \ee

Passing to the mercury atom EDM, the major contribution here
comes from $T$-odd nuclear
forces in $\pi^0$ and $\eta$ couplings to the nucleus, which is
generated by the chromoelectric EDMs of the constituent quarks.
The resulting EDM of the mercury atom is given by ref.\cite{falk}
as \be d_{H_g}=
- e (d^C_d -d^C_u - 0.012 d^C_s) \times 3.2 \times 10^{-2}. \ee
Although the coefficient for the $d^C_s$ is much smaller than
the coefficients of the chromoelectric EDM of the down and up quarks, this
contribution is still important since $d^C_s$ itself
is enhanced by the heavy strange
quark mass and by the relatively large mixing in the second generation.
Recently the mercury EDM has been reconsidered in the light of the QCD sum 
rule calculations, with the result that the coefficients multiplying the 
first generation quarks could be reduced by a 
factor 2.5-3 \cite{pospelov} (see ref.\cite{demir}) for a recent discussion). 
Our study will depend mainly on the strange quark EDM so 
this uncertainty will not effect our conclusions. We
will therefore use the older bound for this study, and comment at the end.

The dominant 1-loop gluino contribution to the EDMs is given by \bea
d^E_{d,u} &=& -\frac{2}{3}~ \frac{\alpha_s}{\pi}~ Q_{d,u}~
\frac{m_{\tilde{g}}}{m_{\tilde{d}}^2}~ \im(\delta^{d,u}_{LR})_{11}~
M_1(x), \label{dEd}\\
d^E_{s} &=& -\frac{2}{3}~ \frac{\alpha_s}{\pi}~ Q_{s}~
\frac{m_{\tilde{g}}}{m_{\tilde{d}}^2}~ \im(\delta^{d}_{LR})_{22}~
M_1(x) , \label{dEs}\\
d^C_{s} &=& \frac{g_s \alpha_s}{4 \pi}~
\frac{m_{\tilde{g}}}{m_{\tilde{d}}^2}~ \im(\delta^{d}_{LR})_{22}~
M_2(x), \label{dCs} \eea where $x=m^2_{\tilde{g}}/m^2_{\tilde{d}}$.
The current experimental bounds using the parton model for the
neutron EDM imply the following constraints on the relevant mass
insertions \cite{abel}: \be \im(\delta^d_{LR})_{11} < 1.9 \times
10^{-6}, ~~~~~ \im(\delta^d_{LR})_{22} < 6.6 \times 10^{-6}, \ee
where to illustrate we have taken $m_{\tilde{d}}\simeq 500$ GeV and
$x=1$. The experimental limit on the mercury EDM leads to a stronger
bound on the imaginary part of $(\delta^d_{LR})_{11}$ and about the
same bound on the imaginary part of $(\delta^d_{LR})_{22}$: \be
\im(\delta^d_{LR})_{11} < 6.7 \times 10^{-8}, ~~~~~
\im(\delta^d_{LR})_{22} < 5.6 \times 10^{-6}. \label{Hgbounds} \ee

As alluded to above, the mass insertion $(\delta^d_{LR})_{22}$ is
more sensitive to the mixing between the second and the third
generation, so the bound on its imaginary part is the relevant one
for our analysis. It is remarkable that the bounds obtained on this
quantity from the mercury EDM and neutron EDM are almost the same,
however, as we emphasized, the computation of the neutron EDM is
more model dependent. Therefore in our analysis we will concentrate
on the constraint obtained from the mercury EDM.

The explicit dependencies of $(\delta^d_{LR})_{22}$ and
$(\delta^d_{RL})_{22}$ on the LL and RR mixing between the second
and the third generations are given by \bea (\delta^d_{LR})_{22} &=&
(\delta^d_{LL})_{23}~ (\delta^d_{LR})_{33}~ (\delta^d_{RR})_{32} +
\left[\delta^d_{RR})_{23}~ (\delta^d_{RL})_{33}~
(\delta^d_{LL})_{32}\right]^*, \eea where $(\delta^d_{LR})_{33}=
(\delta^d_{RL})^*_{33}\sim \frac{m_b (A_b - \mu \tan
\beta)}{m^2_{\tilde{d}}}$. Recall that the EDM is proportional to
the imaginary part of the coefficients of the $d_L^* d_R$ term in the
Lagrangian. In the MSSM the relevant part of the Lagrangian is
given by $\mathcal{L} \sim (Y A - \mu \tan \beta)~ d^*_L d_R ~+~
h.c. $ where $h.c.$ refers to $(Y A - \mu \tan \beta)^* ~d_L d_R^*$.

The EDM imposes stringent constraints on the flavour conserving CP
phases of the $A_b$ and $\mu$ terms. It is reasonable therefore to
assume that these phases are suppressed, in which case
$(\delta^d_{LR})_{33} \sim m_b/m_{\tilde{d}} \sim 10^{-2}$. Also,
due to the hermiticity of the $LL$ and $RR$ sectors of the squark
mass matrix, $(\delta^d_{LL(RR)})_{32} =
(\delta^d_{LL(RR)})^*_{23}$. Thus one finds \bea
(\delta^d_{LR})_{22} &\simeq& 10^{-2} \left[(\delta^d_{LL})_{23}~
(\delta^d_{RR})^*_{23} + \left((\delta^d_{RR})_{23}~
(\delta^d_{LL})^*_{23} \right)^* \right] \eea Furthermore,
$(\delta^d_{LR})_{22}$ can also be expressed as \bea
(\delta^d_{LR})_{22} =(\delta^d_{LL})_{23}~ (\delta^d_{LR})_{32}+
\left[(\delta^d_{RR})_{23}~ (\delta^d_{RL})_{32}\right]^*, \eea
where $(\delta^d_{LR})_{32}= (\delta^d_{RL})^*_{23}$. In the next
section, we will determine the values of $\im (\delta^d_{LR})_{22}$
and $\im (\delta^d_{RL})_{22}$ within the regions of the parameter
space that satisfy the experimental results of $S_{\phi K}$ and
$S_{\eta' K}$. We will show that the EDM of the strange quark allows
the possibility of SUSY models with large LR (RL) mixing only.
%
\section{SUSY contributions to the CP asymmetries $S_{\phi K}$ and $S_{\eta' K}$}
Including the SUSY contribution, the effective Hamiltonian
$H^{\Delta B=1}_{\rm eff}$ for these processes can be expressed
via the Operator Product Expansion (OPE) as \bea H^{\Delta
B=1}_{\rm eff}&=&\left\{ \frac{G_F}{\sqrt{2}} \sum_{p=u,c}
\lambda_p ~\left( C_1 Q_1^p + C_2 Q_2^p + \sum_{i=3}^{10} C_i Q_i
+ C_{7\gamma} Q_{7\gamma} + C_{8g} Q_{8g} \right) + \hc \right\}
\nonumber\\
&+&\left\{Q_i\to \tilde{Q}_i\, ,\, C_i\to \tilde{C}_i\right\} \;,
\label{Heff} \eea where $\lambda_p= V_{pb} V^{\star}_{p s}$, with
$V_{pb}$ the unitary CKM matrix elements satisfying
$\lambda_t+\lambda_u+\lambda_c=0$, and $C_i\equiv C_i(\mu_b)$ are
the Wilson coefficients at low energy scale $\mu_b\simeq m_b$.

As emphasized in refs.\cite{emidio,KK}, the dominant gluino
contributions are due to the QCD penguin diagrams and
chromo-magnetic dipole operators. At the first order in MIA, the
the gluino contributions to the corresponding Wilson coefficients
at the SUSY scale are given by
\begin{eqnarray}
C^{\tilde{g}}_3  &=& -\frac{\alpha_s^2}{2 \sqrt{2}G_F
m_{\tilde{q}}^2} \dd{LL}{23} \left[ -\frac{1}{9} B_1(x) -
\frac{5}{9} B_2(x) -
\frac{1}{18} P_1(x) -\frac{1}{2} P_2(x) \right],\nonumber\\
C^{\tilde{g}}_4  &=& -\frac{\alpha_s^2}{2\sqrt{2} G_F
m_{\tilde{q}}^2} \dd{LL}{23} \left[ -\frac{7}{3} B_1(x) +
\frac{1}{3} B_2(x) + \frac{1}{6} P_1(x)
+\frac{3}{2} P_2(x) \right],\nonumber\\
C^{\tilde{g}}_5  &=& -\frac{\alpha_s^2}{2\sqrt{2}G_F
m_{\tilde{q}}^2} \dd{LL}{23} \left[ \frac{10}{9} B_1(x) +
\frac{1}{18} B_2(x) - \frac{1}{18} P_1(x)
-\frac{1}{2} P_2(x) \right],\nonumber \\
C^{\tilde{g}}_6  &=& -\frac{\alpha_s^2}{2\sqrt{2} G_F
m_{\tilde{q}}^2} \dd{LL}{23} \left[ -\frac{2}{3} B_1(x) +
\frac{7}{6} B_2(x) + \frac{1}{6} P_1(x)
+\frac{3}{2} P_2(x) \right],\nonumber\\
C^{\tilde{g}}_{8g} \!\! &=&\!\!\frac{ \alpha_s \pi} {\sqrt{2}G_F
m_{\tilde{q}}^2}\!\left[ \!\dd{LL}{23}\left( \frac{1}{3} M_3(x)\!
+ \!3 M_4(x)\right)\!+\! \dd{LR}{23}\frac{m_{\tilde{g}}}{m_b}
\left(\!\frac{1}{3} M_1(x) \!+\! 3 M_3(x)\right)\!\right]\!,
\label{Cgluino}
\end{eqnarray}
where $\tilde{C}_{i,8g}$ are obtained from $C_{i,8g}$ by
exchanging $L \leftrightarrow R$ in $(\delta_{AB}^d)_{23}$. It is
clear that the part proportional to LR mass insertions in
$C^{\tilde{g}}_{8g}$ which is enhanced by a factor
$m_{\tilde{g}}/m_b$ would give a dominant contribution. Using the
QCD factorization mechanism to evaluate the matrix elements, the
decay amplitude of $B\to \phi K$ can be presented in terms of the
relevant Wilson coefficients as follows \cite{emidio}:
\begin{equation}
A(B\to \phi K ) = - i \frac{G_F}{\sqrt{2}} m_B^2 F_+^{B\to K}
f_{\phi} \sum_{i=1..10,7\gamma,8g} H_i(\phi) ({\bf C}_i+ {\bf
\tilde{C}}_i),
\end{equation}
where $H_i(\phi)$ are given in Ref.\cite{emidio} and the Wilson
coefficients ${\bf C}_{ i}$ and ${\bf \tilde{C}}_{i}$ are defined
according to the parametrization of the effective Hamiltonian in
Eq.(\ref{Heff})
\begin{equation}
H^{\Delta B=1}_{\rm eff}=\frac{G_F}{\sqrt{2}}\sum_i \left\{ {\bf
C}_{ i} Q_i \, +\, {\bf \tilde{C}}_{i} \tilde{Q}_i\right\}
\label{Heff_NP}
\end{equation}
Therefore, the contributions of the $RR$ and $RL$ terms in
$R_{\phi}$ have the same sign as the $LL$ and $LR$ ones. For
instance, with $m_{\tilde{q}}=m_{\tilde{g}}=500$\ GeV, one obtains
\begin{equation}
R_{\phi} \simeq -0.14~ e^{-i\,0.1}
 (\delta_{LL}^d)_{23}-127~ e^{-i\,0.08}
(\delta_{LR}^d)_{23} - 0.14~ e^{-i\,0.1}
 (\delta_{RR}^d)_{23}- 127~ e^{-i\,0.08}
(\delta_{RL}^d)_{23}. \label{Rgl_Phi}
\end{equation}

From this result, it is clear that the largest SUSY effect is
provided by the gluino contribution to the chromomagnetic operator
which is proportional to $(\delta_{LR}^d)_{23}$. However, the
$b\to s \gamma$ constraints play a crucial role in this case. For
the above SUSY configurations, the $b\to s \gamma$ decay
constrains the possible gluino contributions since it sets
$\vert(\delta_{LR}^d)_{23}\vert< 0.016$. Despite this, on
implementing the bound in Eq.(\ref{Rgl_Phi}), we see that the
gluino contribution (proportional to $(\delta_{LR}^d)_{23}$) is
still able to generate large values for $R_{\phi}$, consequently
driving $S_{\Phi K}$ towards the region of small values.

Although $B \to \phi K$ and $B \to \eta^{\prime} K$ are very
similar processes, the parity of the final states can vary the
result. In  $B \to \phi K$ the contributions from $C_i$ and
$\tilde{C}_i$ to the decay amplitude are identically the same
(with the same sign), whereas in $B \to \ep K$ they have opposite
signs. This can be easily understood by noting that
\begin{equation}
\langle \phi K \vert Q_i \vert B \rangle = \langle \phi K \vert
\tilde{Q}_i \vert B \rangle\, . \label{nfQQt}
\end{equation}
which is due to the invariance of strong interactions under parity
transformations, and to the fact that initial and final states
have the same parity. However, in the case of the
$B \to \ep K$ transition,
where the initial and final states have opposite parity, we have
\begin{equation}
\langle \eta' K  \vert Q_i \vert B \rangle_{QCDF} = - \langle
\eta' K \vert \tilde{Q}_i \vert B \rangle_{QCDF}.
\end{equation}

As a result, the signs of the $C_i$ and $\tilde{C}_i$ in the decay
amplitude are different for $B \to \ep K$, and so the sign of the
$RR$ and $RL$ in $R_{\eta'}$ are different from the sign of $LL$
and $LR$ in contrast with the $R_{\phi}$ case. Using the same SUSY inputs
adopted in Eq.~(\ref{Rgl_Phi}), we have \bea
R_{\eta^{\prime}}\simeq -0.07\, e^{i\,0.24}
 (\delta_{LL}^d)_{23}\,-\,
64(\delta_{LR}^d)_{23}+ 0.07\, e^{i\,0.24}
 (\delta_{RR}^d)_{23}\,+\,
64(\delta_{RL}^d)_{23} \label{Rgl_eta} \eea

Following the parametrization of the SM and SUSY amplitudes in
Ref.\cite{KK}, $S_{\phi K}$ can be written as
\begin{eqnarray}
S_{\phi (\eta') K}=\Frac{\sin 2 \beta +2 R_{\phi (\eta')} \cos
\delta_{12} \sin(\theta_{\phi (\eta')} + 2 \beta) + R_{\phi
(\eta')}^2 \sin (2 \theta_{\phi (\eta')} + 2 \beta)}{1+ 2 R_{\phi
(\eta')} \cos \delta_{12} \cos\theta_{\phi (\eta')} +R_{\phi
(\eta')}^2}~, \label{CPasym}
\end{eqnarray}
where $ R_{\phi}= \vert A^{\susy}/A^{\sm}\vert$, $\theta_{\phi}=
\mathrm{arg}(A^{\susy}/A^{\sm})$, and $\delta_{12}$ is the strong
phase. In order to accommodate the experimental results of
$S_{\phi K}$ and $S_{\eta' K}$ we should have at least one of
the following two scenarios \cite{KK,emidio}: large mixing between the
second and the third generations in $LL$ and $RR$ sectors or large
mixing between the second and the third generations in $LR$ and $RL$
sectors.

As can be seen from Eq.(\ref{CPasym}), the deviation of
$S_{\phi(\eta') K}$ from $\sin 2 \beta$ strongly depends on the
size of $R_{\phi(\eta')}$. The minimum values of $S_{\phi K}$ and
$S_{\eta' K}$ can be obtained by large values of
$\vert(\delta^d_{LL})_{23}\vert \sim \mathcal{O}(1)$,
$\vert(\delta^d_{RR})_{23}\vert \sim \mathcal{O}(1)$ and phases of
$(\delta^d_{LL})_{23}$ and $(\delta^d_{RR})_{23}$ of order one or
$\vert(\delta^d_{LR})_{23}\vert \sim \mathcal{O}(10^{-3})$,
$\vert(\delta^d_{RL})_{23}\vert \sim \mathcal{O}(10^{-3})$ and
phases of $(\delta^d_{LR})_{23}$ and $(\delta^d_{RL})_{23}$ of
order one. It is important to note that to have deviation between
$S_{\phi K}$ and $S_{\eta' K}$, the contributions from
$(\delta^d_{LL})_{23}$ $\left((\delta^d_{LR})_{23}\right)$ and
$(\delta^d_{RR})_{23}$ $\left((\delta^d_{RL})_{23}\right)$ should
be different, so that the gluino contribution to $S_{\phi K}$
becomes larger than its contribution to $S_{\eta' K}$ \cite{KK}.
It is also worth mentioning that, as can be seen from
Eq.(\ref{CPasym}), the effects of LL and RR mixing on
$S_{\phi(\eta') K}$ remain limited compared to the effect of LR
and RL.

\section{EDM constraints on $S_{\phi K}$ and $S_{\eta' K}$}
We now come to the main point of this letter, which is that
such large values for the magnitudes and the phases of
$(\delta^d_{LL})_{23}$ and $(\delta^d_{RR})_{23}$ may
significantly enhance the strange quark EDM thereby
overproducing mercury and possibly neutron EDMs. It is
interesting to ask therefore whether SUSY is still able to accommodate
such large magnitudes and phases, and if so, are they restricted.

As mentioned in the previous section, there are two possible
sources of enhancement:
the first is the combined effect of $(\delta^d_{LL})_{23}$ and
$(\delta^d_{RR})_{23}$, the second source is
either $(\delta_{LL}^d)_{23}$ or $(\delta_{RR}^d)_{23}$
combining with $(\delta_{LR}^d)_{32}$ or
$(\delta_{RL}^d)_{32}$. However, within minimal flavour models such
as minimal supergravity (where the trilinear couplings are
universal), the size of the mass insertions $(\delta^d_{LR})_{23}$
and $(\delta^d_{RL})_{23}$ are of order $10^{-6}$  and $10^{-7}$
respectively. Therefore the imaginary part of the induced mass
insertion $(\delta^d_{LR})_{22}$ can easily be below the bound
obtained from the experimental limit on Hg-EDM. In Fig. 1, we plot
both $S_{\phi K}$, $S_{\eta' K}$ and $d_{Hg}/(d_{Hg})_{Exp}$ as
functions of $\vert (\delta^d_{LL})_{23}\vert$ and $\vert
(\delta^d_{RR})_{23}\vert$. We assume that
$\arg[(\delta^d_{LL})_{23}] \simeq
\arg[(\delta^d_{RR})_{23}]\simeq \pi/2$, in order to enhance their
effects on the CP asymmetries of $B$-decays. We also fixed $m_0
=m_{g}=400$ GeV.

\begin{figure}[t]
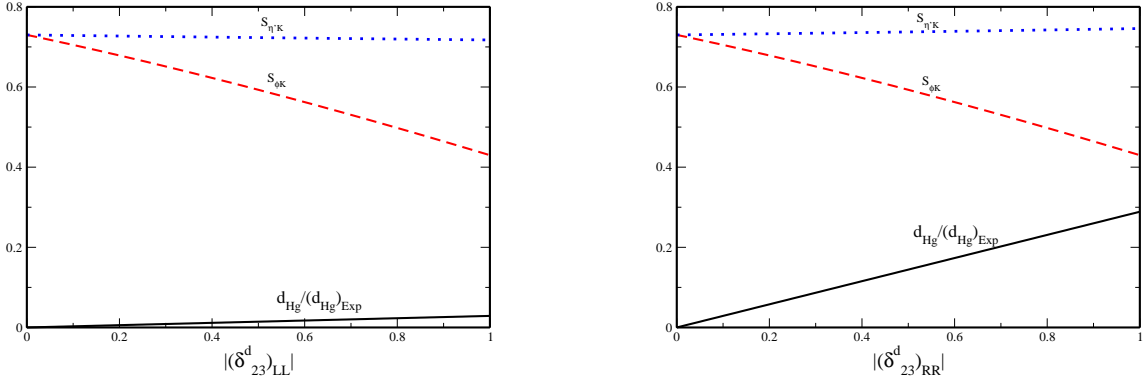

\begin{center}\hspace*{-1cm}
\epsfig{file=LL.eps,width=6.5cm,height=5.cm}\hspace{2cm}
\epsfig{file=RR.eps,width=6.5cm,height=5.cm} \caption{Hg EDM,
$S_{\phi K}$ and $S_{\eta' K}$ as function of the magnitude of
$(\delta^d_{LL})_{23}$ (left plot) and
$(\delta^d_{RR})_{23}$(right plot). The phases of the mass
insertions are assumed to me $\mathcal{O}(\pi/2)$. Also $m_g =
m_{\tilde{q}} = 400$ GeV is used.} \vspace{-0.6cm}
\end{center}
\end{figure}

As can be seen from the figure, in these scenarios the
values of Hg EDM are well below the current experimental limit.
However, we cannot account for the CP asymmetries $S_{\phi K}$
and $S_{\eta' K}$, particularly $S_{\eta' K}$ which has been
the subject of recent measurements by the BaBar collaboration.
In this class of models
with dominant $(\delta^d_{LL})_{23}$ or $(\delta^d_{RR})_{23}$
mass insertions, the value of $S_{\eta' K}$ is close to the SM prediction
of $\sin 2 \beta$. Therefore if the
present $S_{\eta' K}$ result is confirmed, these models will be
disfavoured.

\begin{figure}[t]
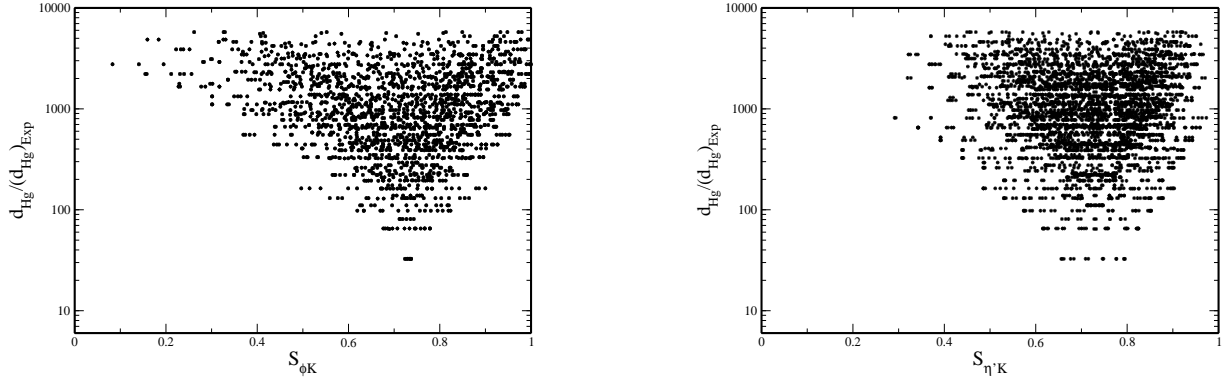

\begin{center}\hspace*{-1cm}
\epsfig{file=LLRR1.eps,width=7.cm,height=5.cm}\hspace{2cm}
\epsfig{file=LLRR2.eps,width=7.cm,height=5.cm} \caption{$S_{\phi
K}$ (left plot) and $S_{\eta' K}$ (right plot) versus the ratio of
the $H_g$ EDM to its experimental value.} \vspace{-0.6cm}
\end{center}
\end{figure}

In Fig. 2, we display scattering plots for $S_{\phi K}$  and
$S_{\eta' K}$ versus the ratio of Hg EDM to its experimental
limit. We set $m_0 = m_g= 400$ GeV. The other relevant parameters
are scanned as follows: $\vert (\delta^d_{LL})_{23} \vert$ varies
from $0$ to $1$, $\arg[(\delta^d_{LL})_{23}]$ and
$\arg[(\delta^d_{RR})_{23}]$ are in the region $\left[-\pi,
\pi\right]$. As can be easily seen from Fig. 2, within the region
of the parameter space where $S_{\phi K}$ and $S_{\eta K}$ fit the
experimental data, the Hg EDM exceeds with many order of
magnitudes its experimental bound. This imposes severe constraints on
this scenario of simultaneous contribution from $LL$ and $RR$
mixing to accommodate both the experimental results of $S_{\phi
K}$ and $S_{\eta K}$. This result is in agreement with that of
of Ref.\cite{yamaguchi}. Returning to the question of the precise numbers 
in the  
bound, it is clear from the figures that even if the strange quark 
contribution to the mercury EDM were reduced by
a whole order of magnitude (rather than the factor 2.5-3 reduction 
implied for the first generation
contributions to the Hg-EDM from the sum-rule calculations)
this conclusion is unchanged.

Therefore, we may safely conclude 
that SUSY models with dominant $LL$ and/or $RR$ large mixing between second
and third generations will be ruled out if the experimental results
of $S_{\phi K}$, $S_{\eta K}$ are confirmed.
SUSY models with dominant $LR$ and/or $RL$
mixing via non-universal $A$-terms, seem to be the
simplest way to account for CP asymmetry $S_{\phi K}$ and $S_{\eta K}$
without conflicting with EDMs or any other experimental results.

Before we conclude, we give a quantitative prediction for the
Hg-EDM due to the effect of large $(\delta^d_{LR})_{23}$. As shown
in Ref.\cite{emidio}, in order to accommodate the experimental
result of the CP asymmetry $\vert (\delta^d_{LR})_{23}\vert$
should be of order $10^{-3}$ and $\arg[ (\delta^d_{LR})_{23}]\sim
\pi/3$. These values lead to $S_{\phi K} \simeq 0.2$. Assuming
(the minimal assumption) that the soft scalar masses are universal
at the SUSY breaking scale, one finds that at the electroweak
scale $(\delta^d_{LL})_{23}$ is of order $10^{-3}$ and
$(\delta^d_{RR})_{23} \sim 10^{-6}$. Hence one finds that
$(\delta^d_{LR})_{22} \lsim 10^{-6}$ which implies that $d_{Hg}
\sim 0.2 (d_{Hg})_{Exp}$. We find it intriguing that the Hg EDM
experiment is so close to testing CP violation in the flavour
changing sector.

\subsection*{Acknowledgements}

We would like to thank Emidio Gabrielli and Oleg Lebedev for useful communications. This work was partially supported by a NATO Collaborative Linkage Grant.


\end{document}